# Seismic Crystals And Earthquake Shield Application

B. Baykant ALAGÖZ[*]       Serkan ALAGÖZ

**Abstract:** *We theoretically demonstrate that earthquake shield made of seismic crystal can damp down surface waves, which are the most destructive type for constructions. In the paper, seismic crystal is introduced in aspect of band gaps (Stop band) and some design concepts for earthquake and tsunami shielding were discussed in theoretical manner. We observed in our FDTD based 2D elastic wave simulations that proposed earthquake shield could provide about 0.5 reductions in magnitude of surface wave on the Richter scale. This reduction rate in magnitude can considerably reduce destructions in the case of earthquake.*

**Keyword:** Earthquake shield, seismic crystals, band gap
**PACS:** 46.40.-f, 91.30.-f

## Introduction:

From researches, done for photonic and sonic crystals, band gaps were observed in the band structures of crystals with various lattice geometries.[1-5] Since, crystal structures exhibit very low transmission for incoming wave in frequencies coinciding band gaps, it was referred as stop band in some studies. Band gaps occur when homogenous wave propagation medium turned into an inhomogeneous condition by periodical flawing of scattering material. This periodical flaw in medium leads frequency selectivity in transmission of the wave and results transmission bands, permitting wave propagation through the crystal, and band gaps, prohibiting wave propagation through the crystal. Practically, band gaps properties of sonic crystal structures were used in application of reduction of vibrations or sound wave. In this study, we consider crystal structures for suspension of destructive seismic waves from an earthquake in a theoretical manner. Accordingly, we referred these crystals as *Seismic Crystal* and application of isolating a region from destructive seismic wave as *Earthquake Shielding*.

Since difficulties in experimental study of earthquakes, we applied to Finite Difference Time Domain (FDTD) simulations, which was the most common numerical analysis technique used for elastic or acoustic waves in inhomogeneous materials.[1,3] Destructive surface waves travels as compression and decompression of granular structure. In order to model this type of surface wave propagation mathematically, we used elastic wave equation dependent on parameters of particle velocity and pressure. This wave equation was solved for two-dimensional simulation of seismic waves by FDTD method. In our earthquake simulations, elastic waves with very low frequencies (1-10 Hz) were produced and propagated through seismic crystals made of ground and hole. In the simulation of earthquake, magnitude of particle displacement were recorded and maximum particle displacement maps scaled to Richter scale (Vibration Map) were presented to demonstrate efficiency of earthquake shielding. Wavelength response of seismic crystals was obtained from FDTD simulations and they were used to determine lattice constant parameter in designing of seismic crystal for earthquake shielding.

---

- B. Baykant Alagöz – Inonu University, department of Electric-Electronics, Turkey (alagozb@inonu.edu.tr)
- Serkan Alagöz – Inonu University, department of Physics, Turkey (sealagoz@inonu.edu.tr)

## Basics Of Crystals And Band Gaps:

Crystal structures consist of periodic scatters arrays, which are mostly in cylindrical shape in literature, and a host material. In this inhomogeneous medium, propagation of waves will be modulated by the periodic structures of scatters and dispersion of waves will no longer behave as in homogenous medium, transmission bands and band gaps appears in band structure characteristics. To form a periodic inhomogeneous medium, photonic crystals are composed of periodically modulated dielectric materials and sonic crystals are mostly composed of periodic array of solid rods in the air. For the seismic crystals, we preferred periodic holes as scatters and ground as host material. Our main consideration in material selection of seismic crystal is to construction simplification. Proposed seismic crystal composed of periodic hole arrays in the ground would be rather easy to build when compared other options, due to its enormous size discussed in next sections. In the Figure 1, crystal structure for 1D and 2D and its relevant design parameter are represented, basically. In this paper, we were addressed 2D crystals. Distance between centers of cylindrical scattering materials is called as lattice constant and usually denoted by $a$ and radius of cylindrical scattering material is denoted by $r$ in the literature. Filing fraction ($f$) was defined as ratio of scatter area to area of unit cell and it is expressed as $f = \frac{2\pi}{\sqrt{3}}(r/a)^2$ for circular scatters. It was taken account as an important parameter in the band gap studies.

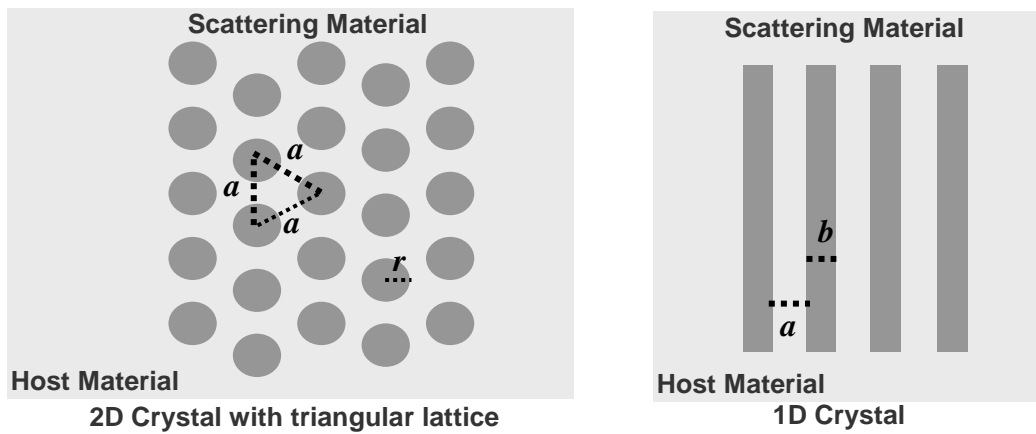

**Figure 1**. Representation of 2D and 1D crystal structures

Band structure of a crystal was presented in the Figure 2 to illustrate band gaps and transmission bands. Band structure and equifrequency surface characteristic of a crystal is commonly computed by Plane Wave Expansion (PWE) method. Details of PWE method was addressed in many study at some decree.[1,6] In this paper, we do not give details about PWE methods. Rather, we give some explanations about transmission characteristic attached to band structure characteristic seen in Figure 2. In the figure, three-transmission band and possible band gaps were representatively drawn together with transmission ratio. In acoustic wave literature, band gaps were shown to provide about 30 dB reductions in transmission [1] and it opens up for a filling fraction, that is $f > 0.08$. Width of band gaps tends to increase with the filling fraction. [5]

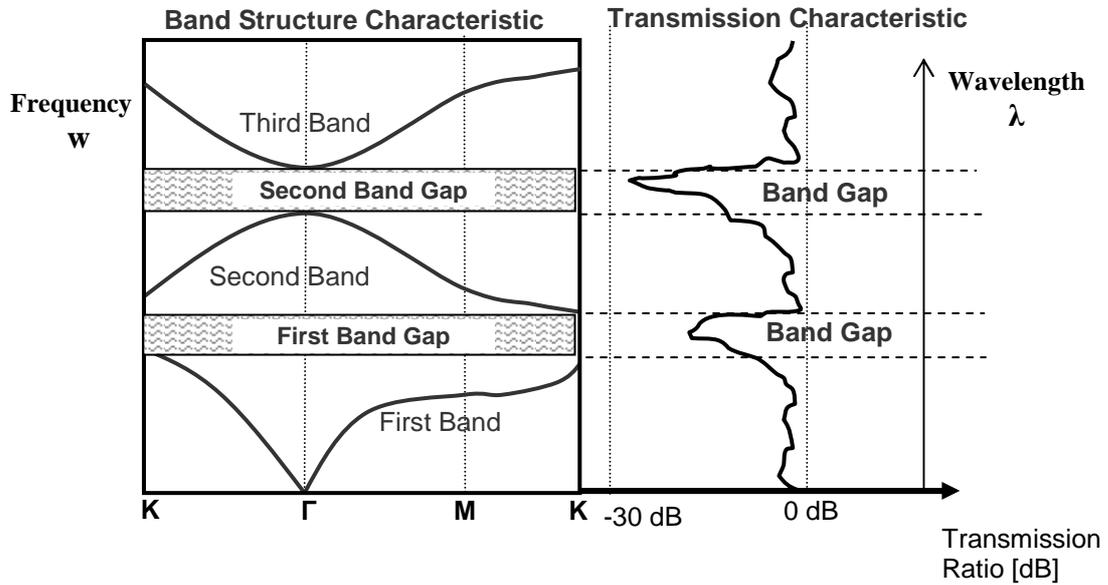

**Figure 2**. An example of band structure and transmission characteristic

## Numerical Analysis By FDTD:

For the numerical model of seismic wave propagation, ground was assumed being made of very tiny granular structures, which is named as particles. Particle displacement in short distance is assumed to result compression and decompression fields on ground. In compressed regions, pressure will be much higher than the pressure of decompressed region as represented in Figure 3.

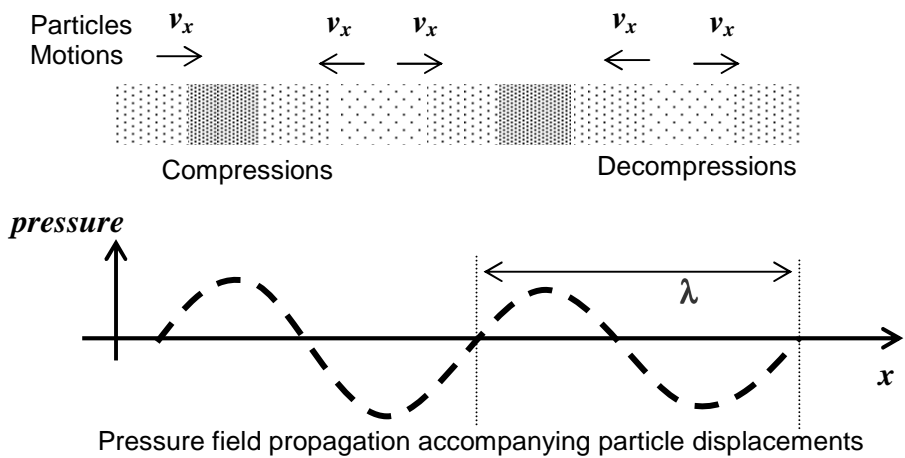

**Figure 3**. Particle displacements and pressure field in one dimension

These differences in pressure fields enforce to particles move in direction to reduce difference in pressure fields and these motions of particles continue until reaching an equilibrium. Energy of seismic wave delivered to system by mean of those particles displacements. This mechanism is well defined by following elastic wave equations,

$$\frac{\partial v}{\partial u} = -\underline{\rho}.\nabla p \quad (1)$$

$$\frac{\partial p}{\partial u} = -\overline{K}.\nabla.v \quad (2)$$

where , $v$ denotes normalized particle velocity and $p$ is the pressure. The material parameters for inhomogeneous structures are described by the normalized medium density, $\underline{\rho} = \rho_o/\rho$, and the normalized bulk modulus, $\overline{K} = K/K_0$. $\rho_o$ is density of host material and $K_o$ is bulk modules of host material. Equations (1) and (2) were proposed for numerical analysis of acoustic wave in literature.[1] In our application, density of host material, that is ground, is much larger than the scatters, which is air in hole and accordingly elastic wave velocity in ground will be 8-10 times larger than speed of sound, which is roughly 340 m/sn.

For the FDTD simulation of seismic waves, we set wave velocity for host material ($c$) to 3000 m/sn. For the scattering material, normalized medium density $\underline{\rho}$ was set to 10 and normalized bulk modulus $\overline{K}$ to 0.1. Spatial differences $\Delta x$ and $\Delta y$ of FDTD simulation were set to 40 $m$ and $\Delta u$ time difference for FDTD simulation was set to $4.66 \cdot 10^{-3}$ second.

Swell of the ground level will be caused by compressed and decompressed fields of ground as illustrated in Figure 4. Compression of particles leads rising of ground level as a result of weight of air and granules body be lower than the force of compression. Since weight of air and granules body is higher than the force of compression (*F*), collapse of ground level will be seen where decompressed field took effect. Maximum particle displacements in direction of z-axis, denoted by $h$, can be simply estimated depending onto pressure as following,

$$h = \alpha \cdot p \quad (3)$$

$\alpha$ is coefficient to rise.

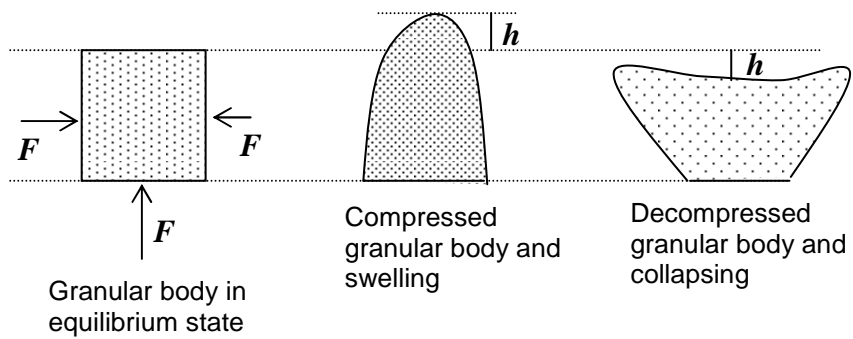

**Figure 4**. Ground swelling by pressure variation

## Earthquake Simulation Results:

In this section, FDTD based simulation results of earthquake shielding are presented respected to Richter scale, which is logarithm of maximum displacement in micron ($M = \log \Delta x_{micron}$). In the simulation, rectangular sheet representing host material was stimulated by a signal of pressure field ($p$) containing multi-frequency components seen in Figure 5. This signal generated by following function,

$$Ps(n) = \sin(\pi \cdot n) + \sin(2\pi \cdot n) + \sin(4\pi \cdot n) + \sin(6\pi \cdot n) + \sin(8\pi \cdot n) + \sin(10\pi \cdot n) + \sin(20\pi \cdot n)$$

Here, $n$ is simulation time index. Simulation time will be $u = n \cdot \Delta u$, $\Delta u$ time difference for FDTD simulation was set to $4.66 \cdot 10^{-3}$ second.

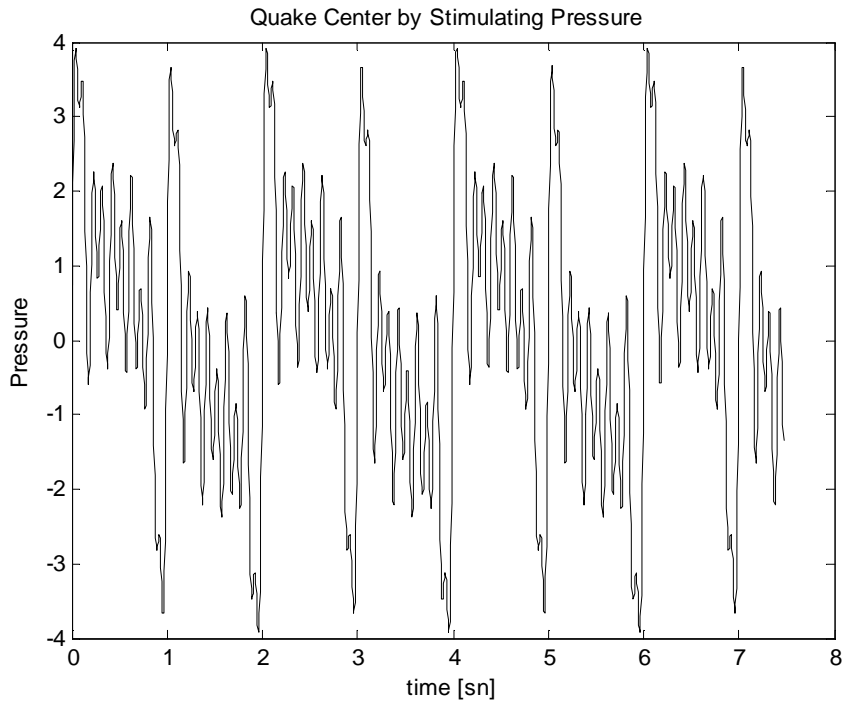

**Figure 5**. Pressure stimulation for earthquake center

In the Figure 6 and 7, pressure wave ($p$) pattern and vibration map in Richter scale were given to make apparent seismic wave reduction by mean of earthquake shielding. Pressure wave pattern ($p$) is also normalized image of swelling, which is ($\frac{h}{\alpha}$) according to equation (3). Vibration map in Richter scale is composed of norm of maximum instant velocity vector parallel to ground (Rectangular sheet in size of *40x40* km$^2$). It is produced by following formula,

$$V_{\max}(i,j) = \max\left(\sqrt{v_x^{\,2}(i,j,n) + v_y^{\,2}(i,j,n)}\right) \quad (4)$$

for $n \in [0, n_{sim}]$. In the formula, $n_{sim}$ determines simulation duration. $v_x(i,j,n)$ and $v_y(i,j,n)$ are component of velocity vector at point $(i, j)$ (See Figure 8) and simulation time at $n$.

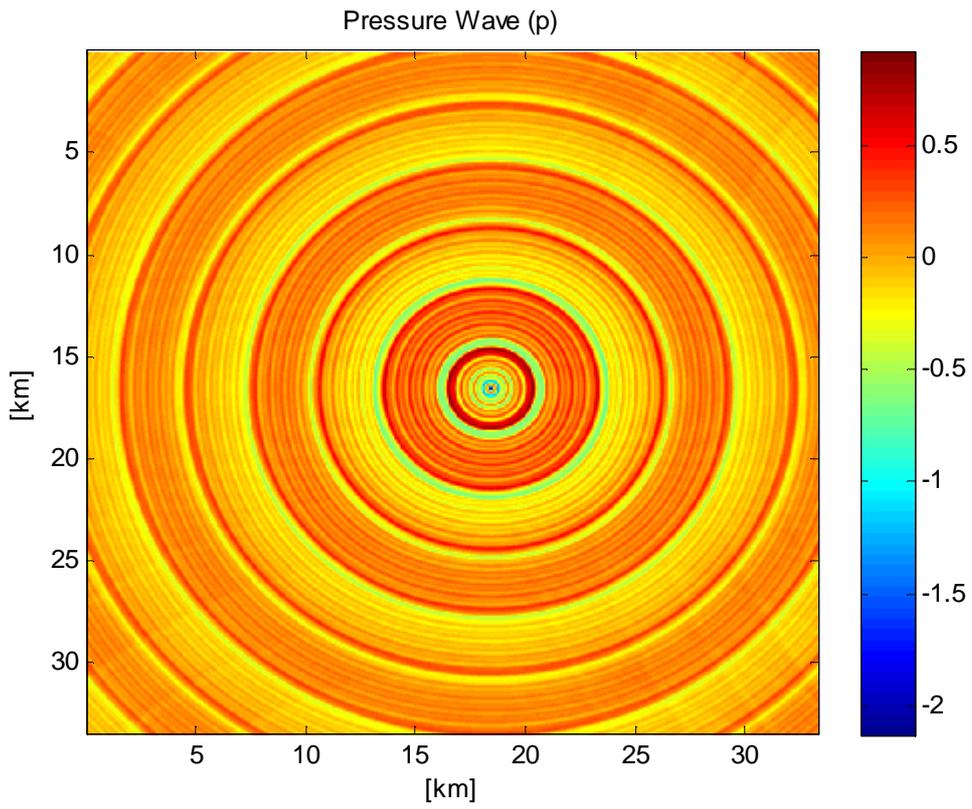

**a.** Pressure wave without seismic crystal

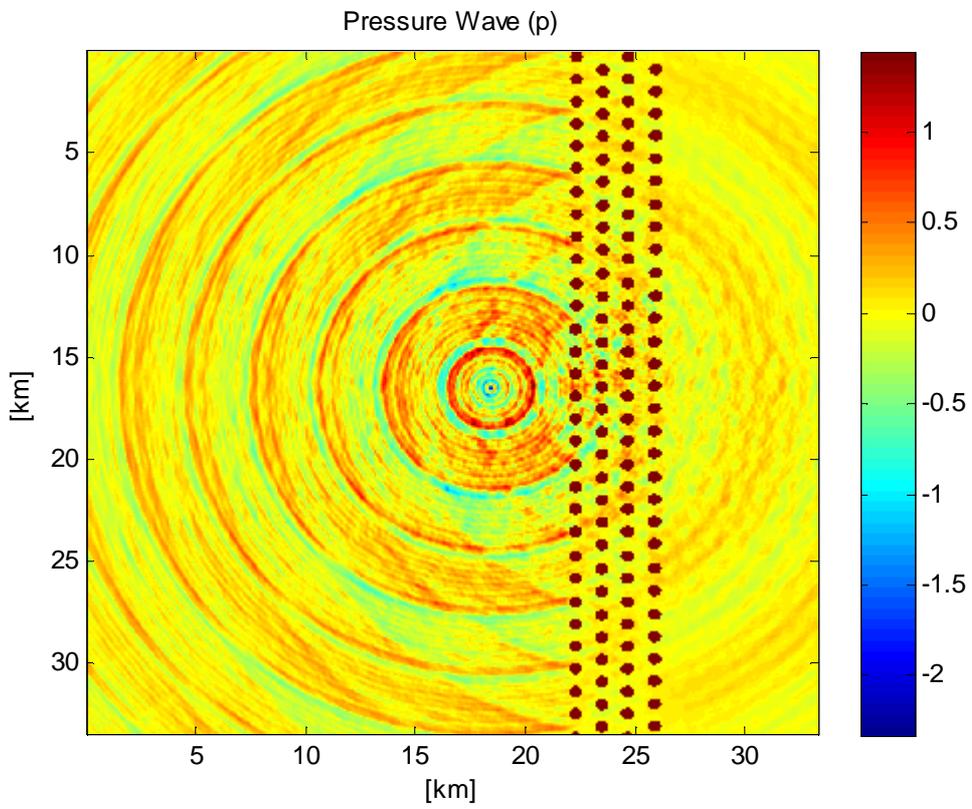

**b.** Pressure wave with seismic crystal triangular lattice configuration

**Figure 6**. Pressure wave (p) patterns

headerpubCEJP-D-09-00056

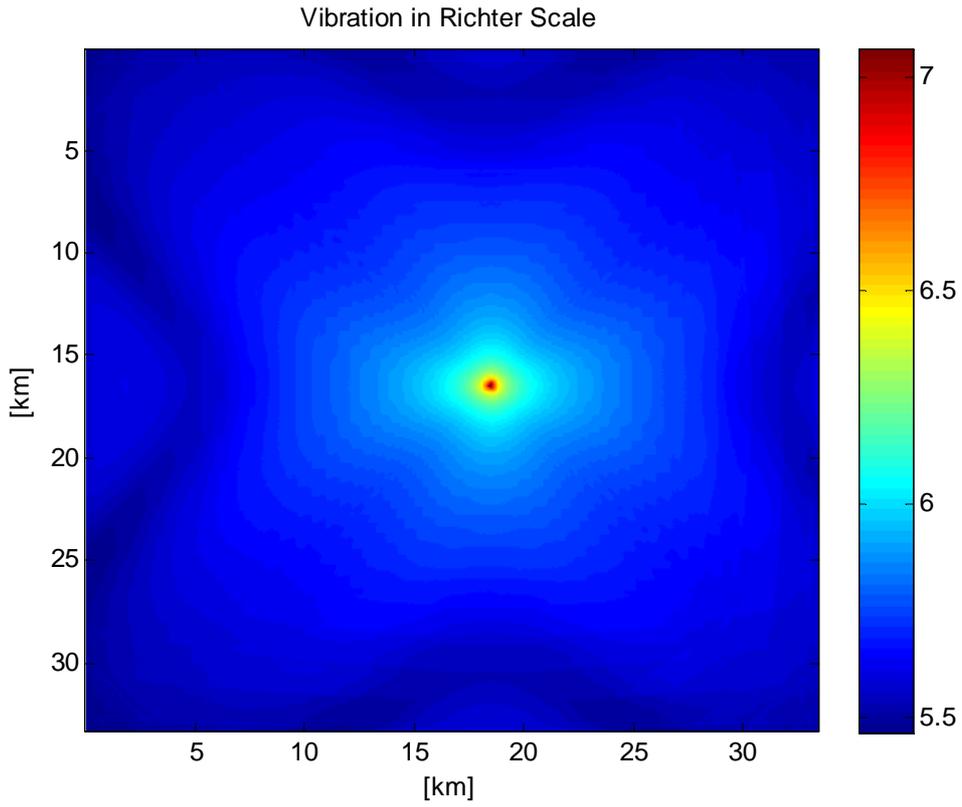

**a.** Vibration in Richter scale without seismic crystal

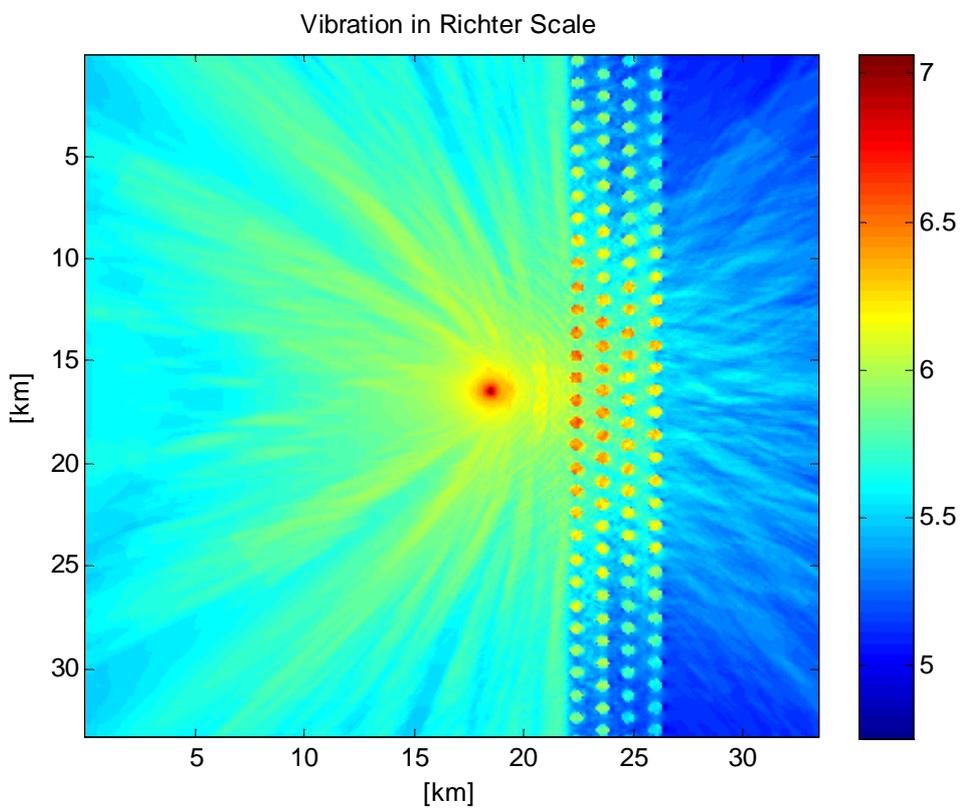

**b.** Vibration in Richter scale with seismic crystal with triangular lattice configuration

**Figure 7**. Vibration in Richter scale

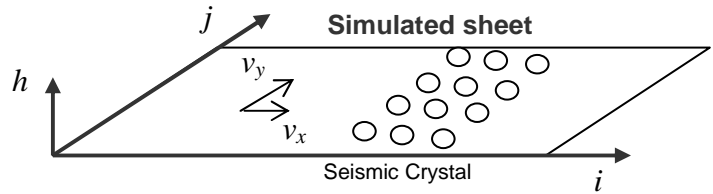

**Figure 8**. Pressure wave (p) pattern and vibration in Richter scale

In the Figure 6, one can see that seismic crystal smoothed sharp pressure changes that can be destructive for constructions. Because, force faced in sharp pressure variations will be much higher than slower pressure variation. For the Figure 7, up to 0.5 dB reductions in vibration can be seen next to seismic crystal when compared simulation results without seismic crystals. In the Figure 9, simulation results for seismic crystal with honeycomb geometry were presented for comparison proposes.

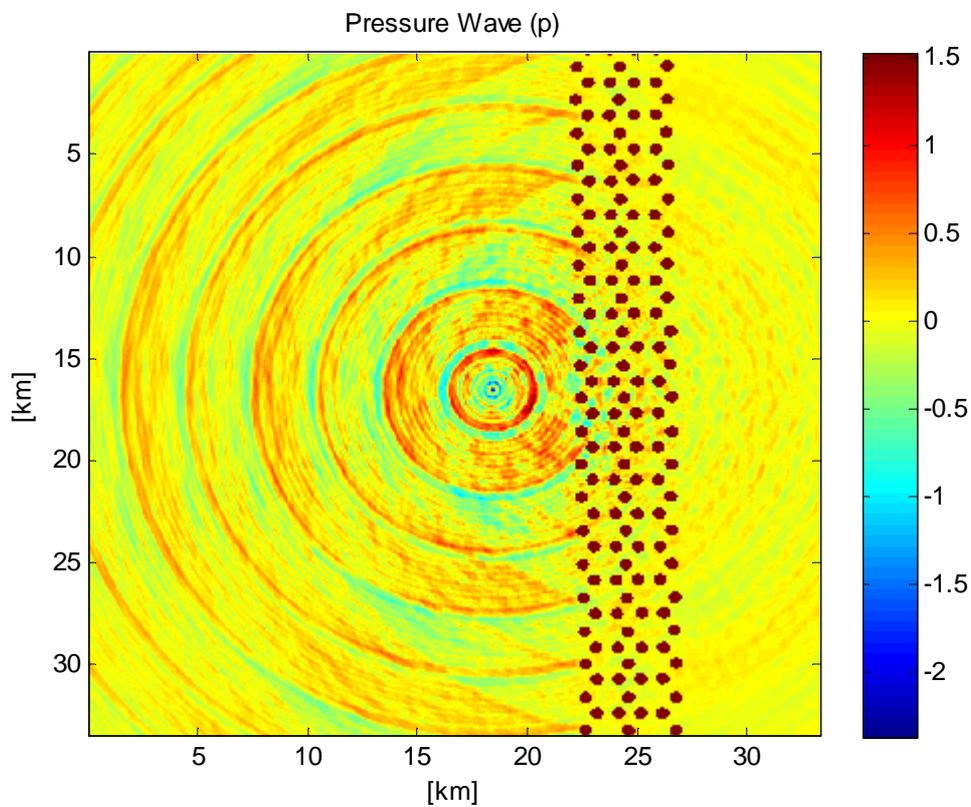

**a.** Pressure wave with seismic crystal honeycomb lattice configuration

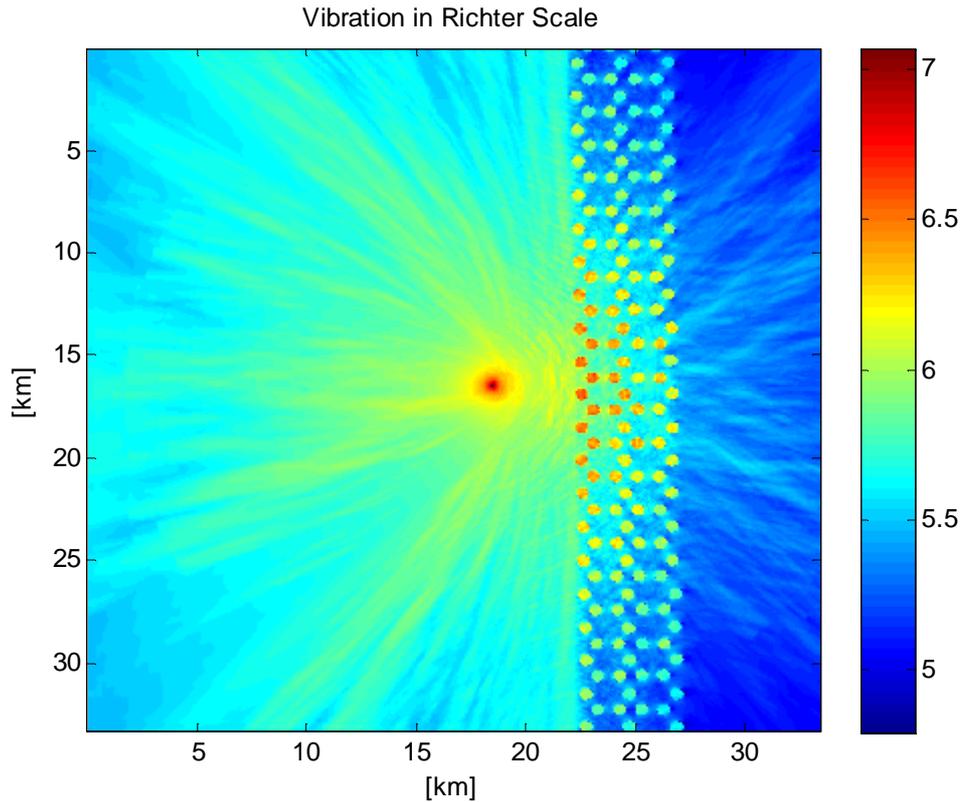

**b.** Vibration in Richter scale with seismic crystal with honeycomb lattice configuration

**Figure 9**. Pressure wave (p) pattern and vibration in Richter scale

When wavelength is much greater than lattice constant, wave can propagate on crystal as if it is in homogenous material. First band gap shows itself when wavelength approximates lattice constant. Seismic wave contains multi-frequency components in frequency spectrum. Lattice constant should be selected respect to $\lambda_{max}$ center wavelength (See Figure 13.a), which is maximum amplitude in spectrum of vibration. Lattice constant selection criteria will be obtained as $a = \dfrac{4 \cdot \lambda_{max}}{9}$ in the "Wavelength Response Of Seismic Crystal" section.

Width of band gap increase by filling fraction ($f$). Unfortunately, higher filling fraction for ground makes crystal more fragile against seismic waves. In order to obtain wider band gaps, cascaded crystals with various lattices constant can be better way. [5] Surface waves has very low frequency ranges between 1 Hz to 10 Hz and wavelengths of 1500-5000 m.

Simulation results for some different seismic crystal types in term of scatter geometry and lattice configuration can be found in the next figures. Crescent and elliptic scatters are seen to have better performance of isolation for seismic waves. Up to 0.8 dB reduction can be seen in vibration map of elliptic scatters. One of disadvantage of circular scatters is that, crystals with circular scatters may cause wave focusing phenomenon which was called as super lens. [7] This phenomenon was observed at some frequency bands where, crystal has negative effective refraction index (ERI). Wave focusing must be avoided since vibration will be amplified at focal point and regions near to it. Seismic wave will possibly be very destructive around focal point. Circler scatter shouldn't be preferred in earthquake shielding due to risk of wave focusing.

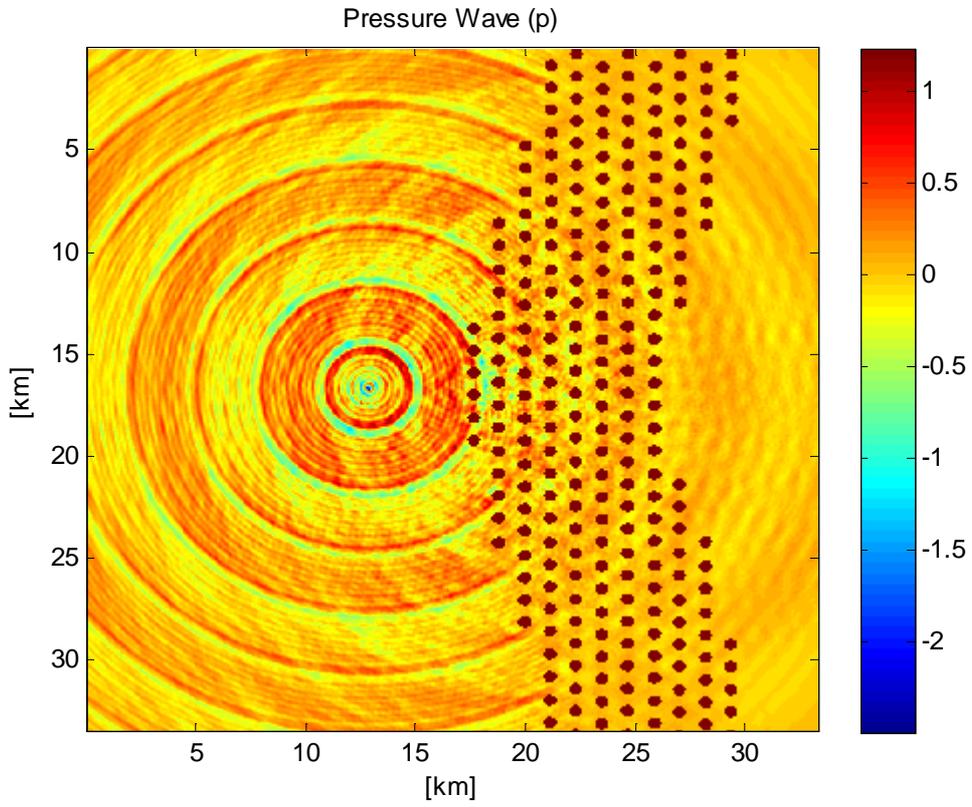

**a.** Circler scatters with triangular lattice configuration (Thick and curved crystal)

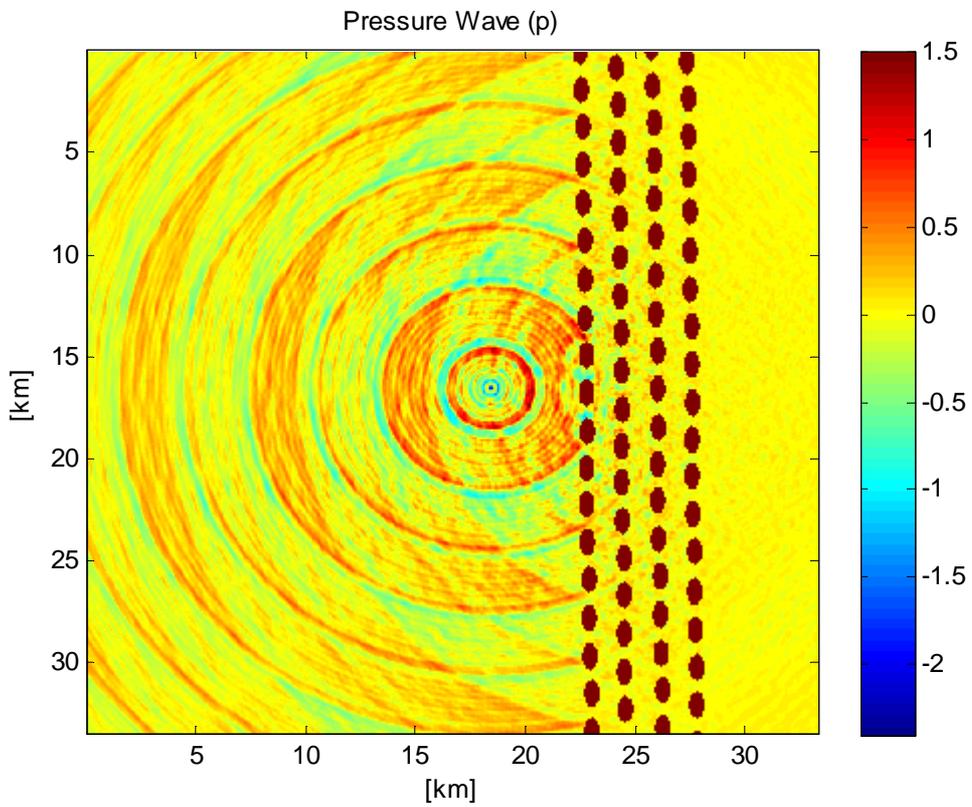

**b.** Elliptic scatters [8] with triangular lattice configuration

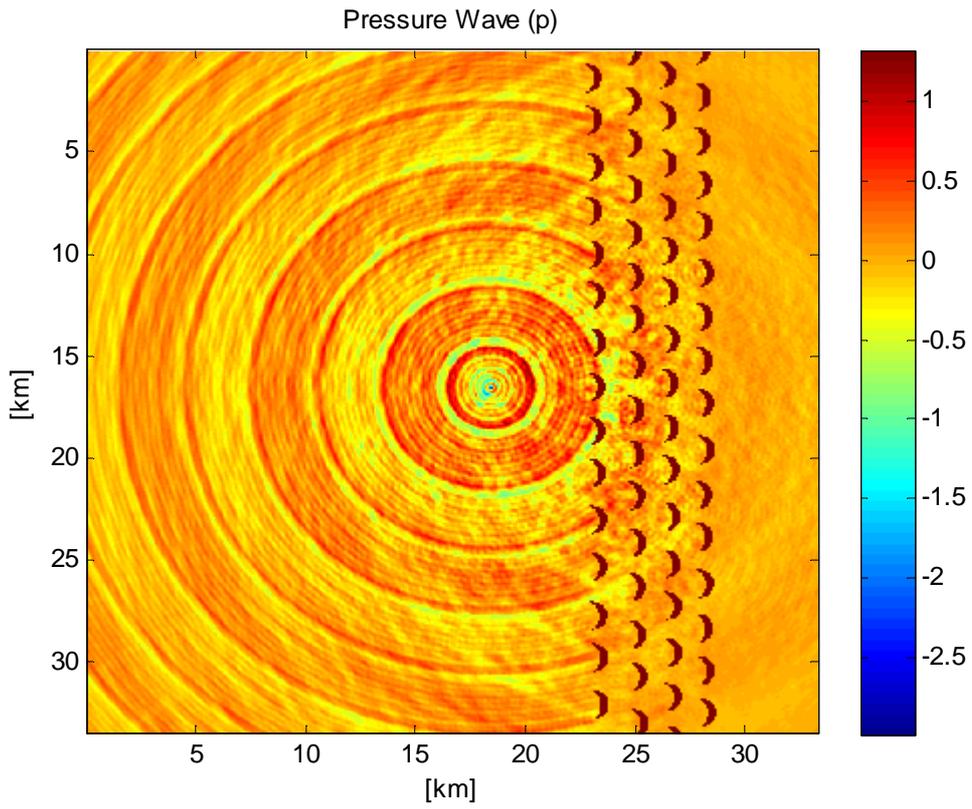

**c.** Crescent scatters with triangular lattice configuration

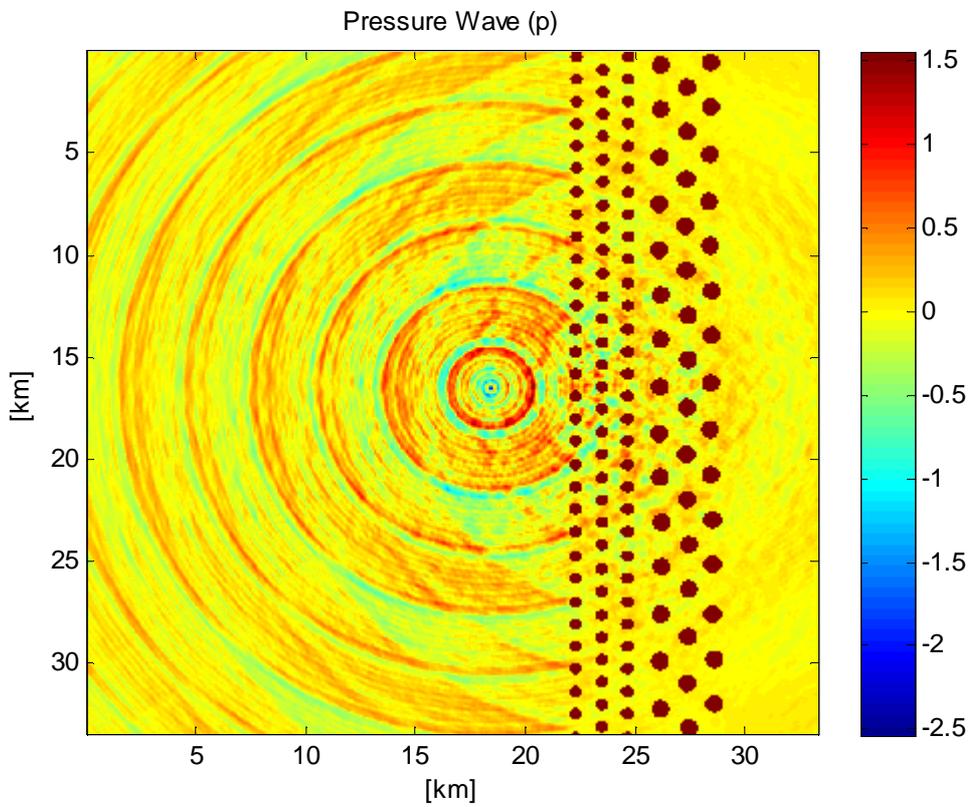

**d.** Cascaded crystal with triangular lattice configuration

**Figure 10**. Pressure wave (p) patterns for various seismic crystals

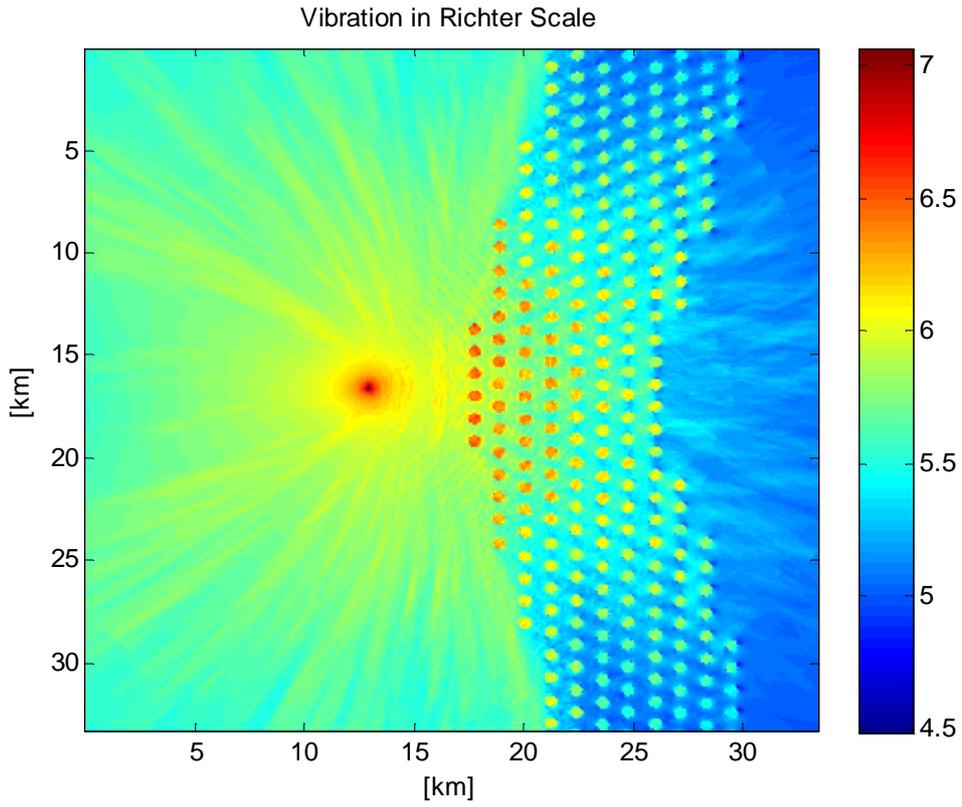

**a.** Circler scatters with triangular lattice configuration

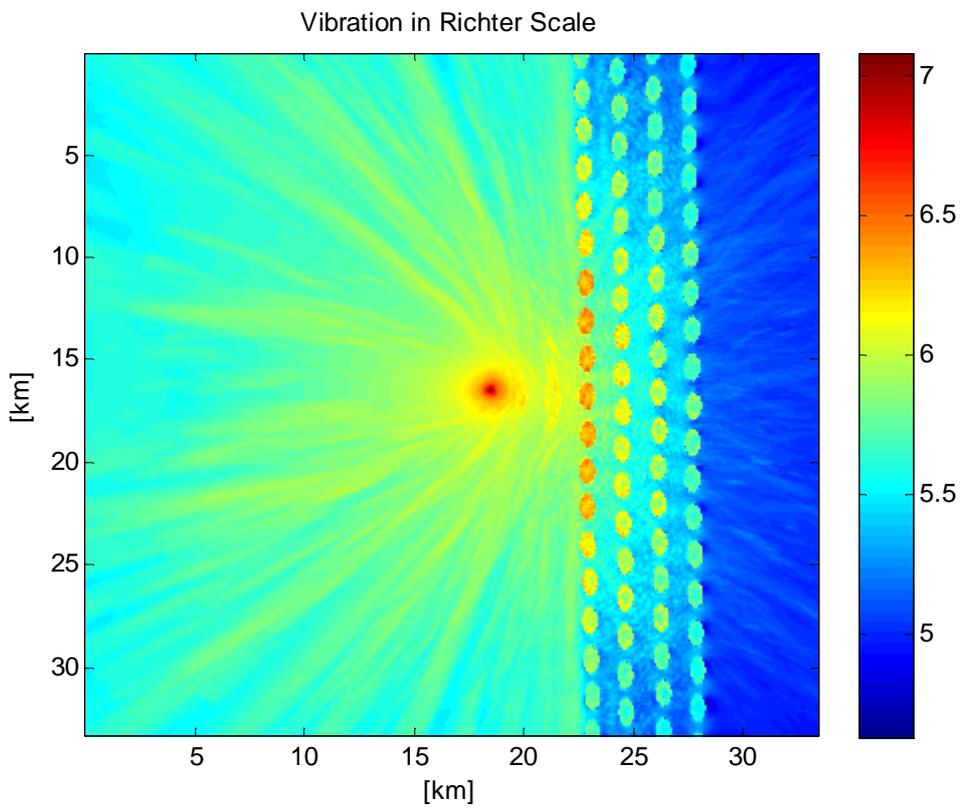

**b.** Elliptic scatters [8] with triangular lattice configuration

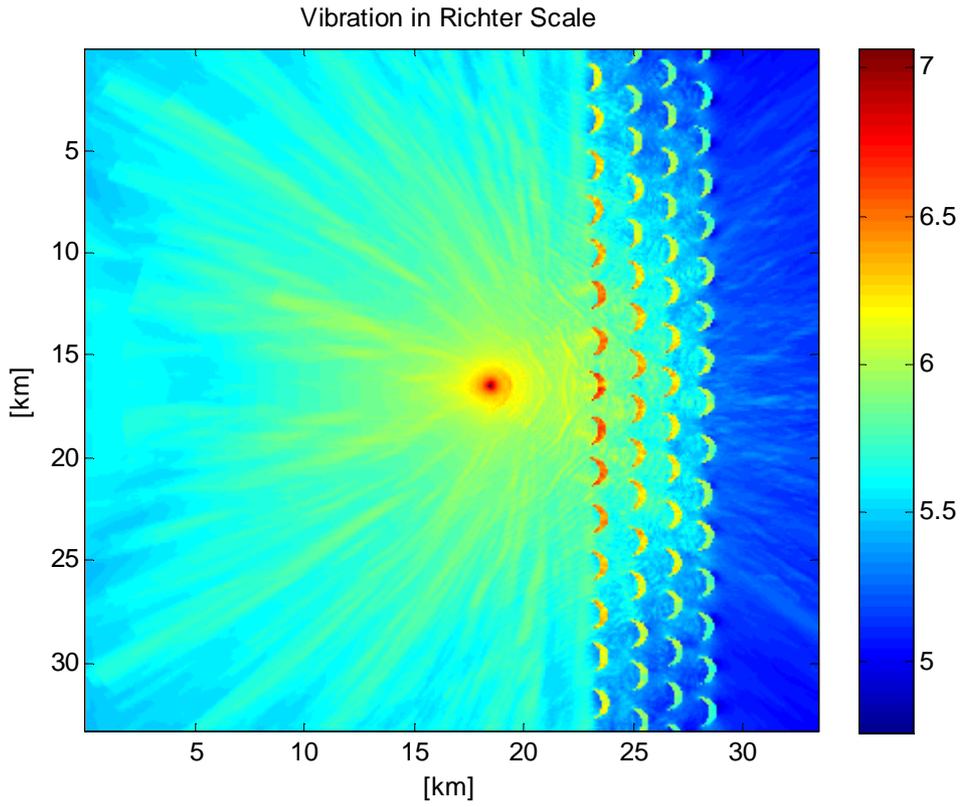

**c.** Crescent scatters with triangular lattice configuration

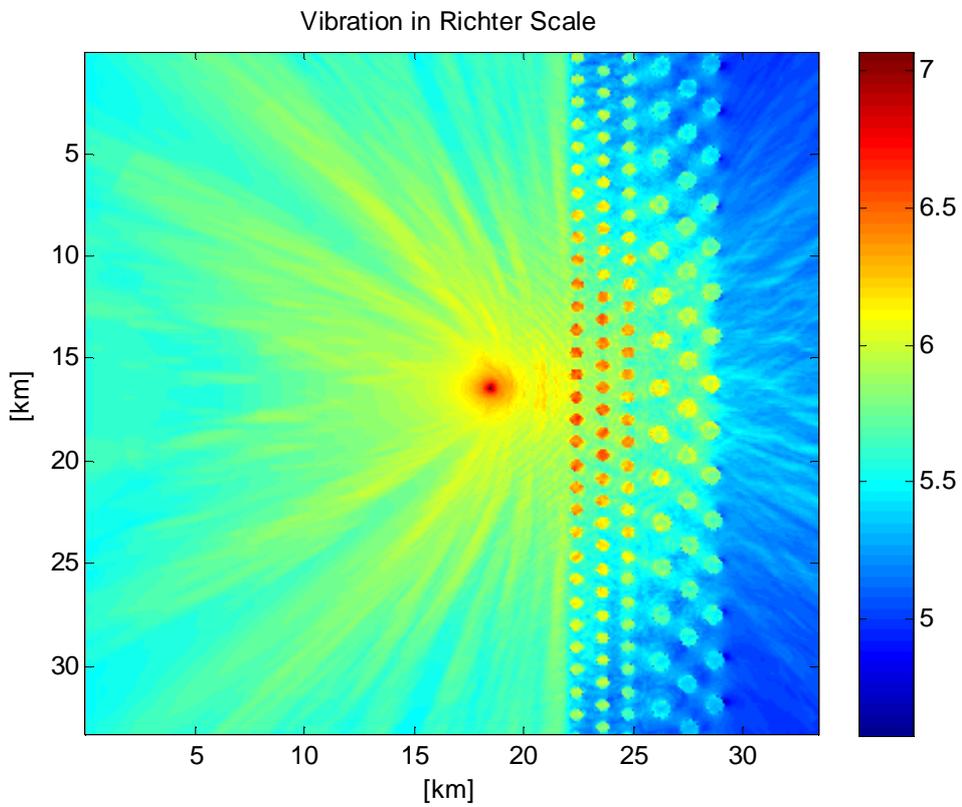

**d.** Cascaded crystal with triangular lattice configuration

**Figure 11**. Vibration in Richter scale for various seismic crystals

Seismic waves, which have very high energy, will inevitably demolish seismic crystals in some decrees. Damages on crystal will reduce its isolation property. At this points, earthquake duration gains importance for shielding property taking effect. When earthquake continues long time, earthquake shield may be completely smashed and it may loose isolation property. Therefore, strength of crystal is an important design consideration limiting filling fraction and width of band gaps.

**Maximum Acceleration Map From FDTD Simulations:**

Maps of maximum particle acceleration give idea about destructive force applied by seismic waves. Maximum particle acceleration in discrete form can be calculated in FDTD simulations by following formula,

$$A_{max}(i,j) = \max\left(\left|\sqrt{v_x^2(i,j,n) + v_y^2(i,j,n)} - \sqrt{v_x^2(i,j,n-1) + v_y^2(i,j,n-1)}\right|\right) \quad (5)$$

for $n \in [0, n_{sim}]$

Force applied on particles can be expressed by,

$$F(i,j) = \rho(i,j) \cdot U_v \cdot A_{max}(i,j) \quad (6)$$

$\rho(i,j)$ is density at point $(i,j)$ and $U_v$ is unit area represented by a points of simulated plane. Maximum particle acceleration and maximum force distribution maps obtained by equation (5) and (6) were illustrated for elliptic scatter geometry at Figure 12. Reduction of the force by crystal structure is apparently seen as darkening region on the right side of crystal.

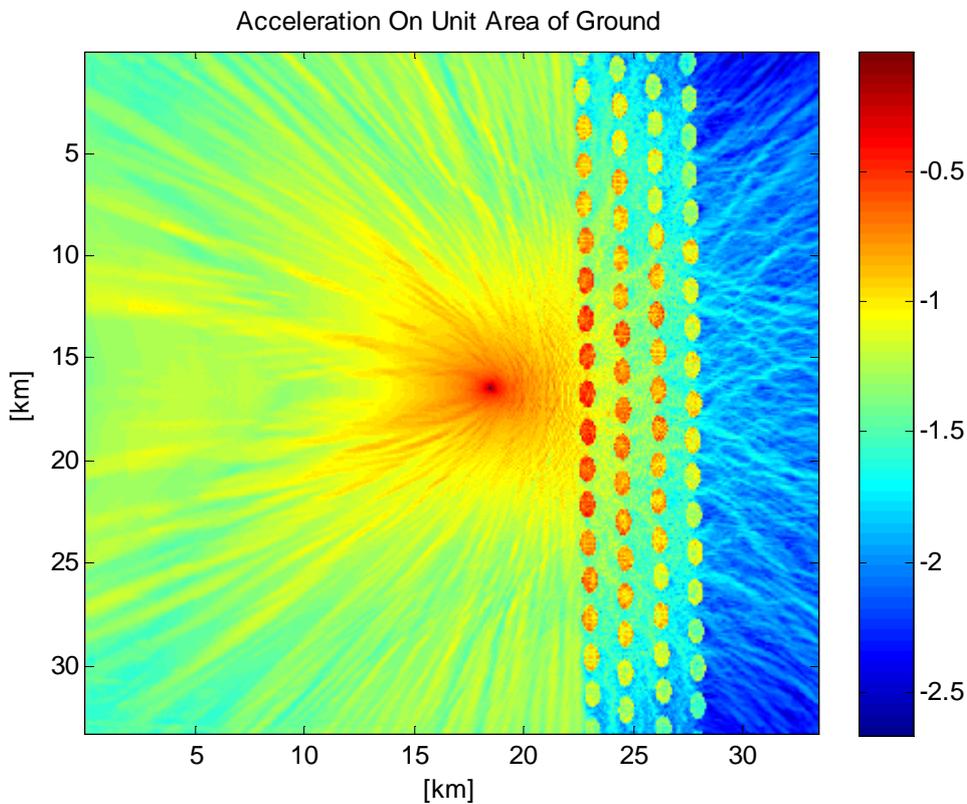

**a.** Maximum particle acceleration map on the logarithmic scale

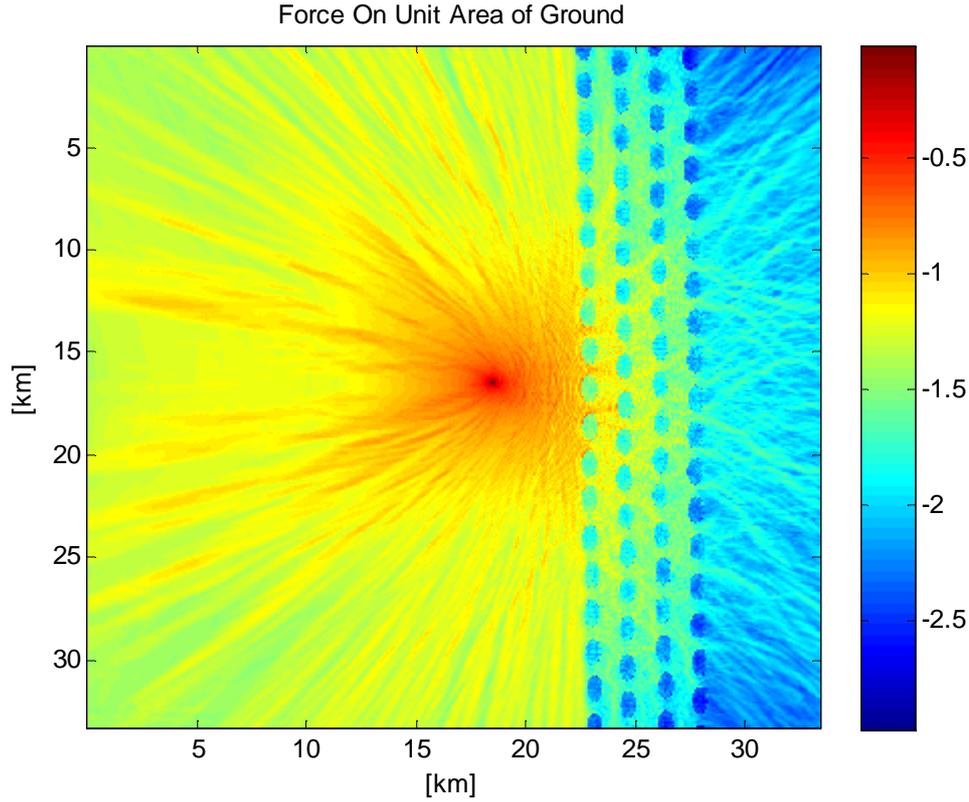

**b.** Maximum force distribution maps on the logarithmic scale

**Figure 12.** Maximum particle acceleration and Maximum force distribution maps.

## Wavelength Response Of Seismic Crystal:

In this section, wavelength response of seismic crystal with various lattice configuration and scatter shape are investigated to figure out convenient lattice constant selection criteria depending on center wavelength $\lambda_{max}$ of vibration signal. For this proposes, we recorded pressure values ($p_{0,\lambda}(n)$, $p_{1,\lambda}(n)$) in FDTD simulation done by monochromatic point source for each wavelength. $p_{1,\lambda}$ holds pressure data from shielded point. $p_{0,\lambda}$ holds pressure data at the same point with $p_{1,\lambda}$ in absence of shielding. Reduction rate of shielding ($R(\lambda)$) for each corresponding wavelength was calculated by maximum pressure values as seen following formulas,

$$R(\lambda) = \frac{\max(p_{1,\lambda})}{\max(p_{0,\lambda})} \qquad (7)$$

Obviously, best damping in vibration signal will be obtained, if $\lambda_{max}$ coincides in middle of the band, where $R(\lambda)$ has a minimum. In such condition, resulting spectrum of vibration after filtering by the seismic crystal will be similar to one given at right of Figure 13.a. Result of these filtering process is displayed in vibration maps and pressure wave patterns at previous sections.

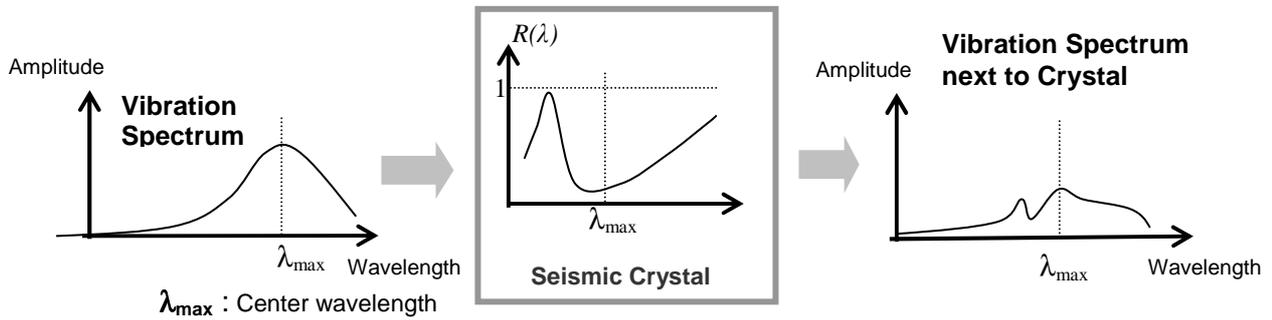

**a.** Vibration spectrum, reduction by seismic crystal and vibration spectrum of reduced vibration

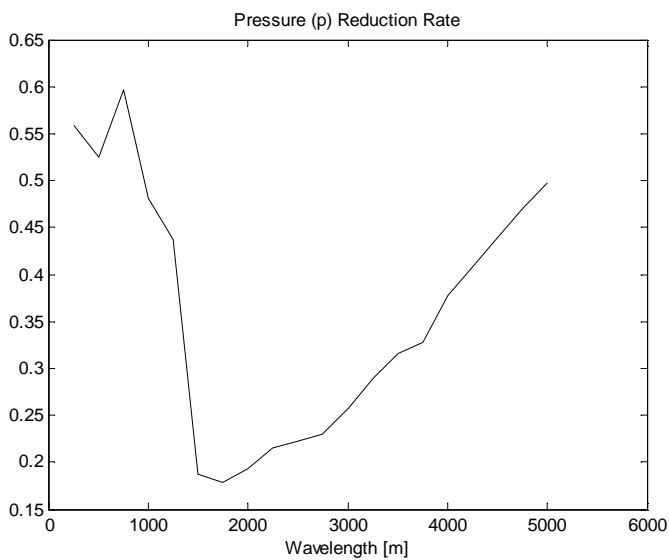 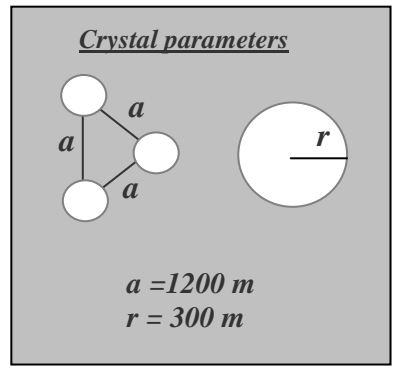

**b.** Reduction rate for triangular lattice configuration and circular scatter

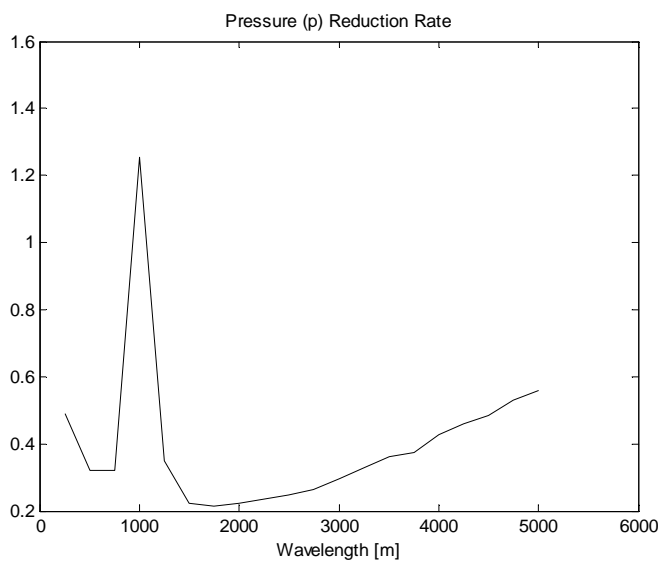 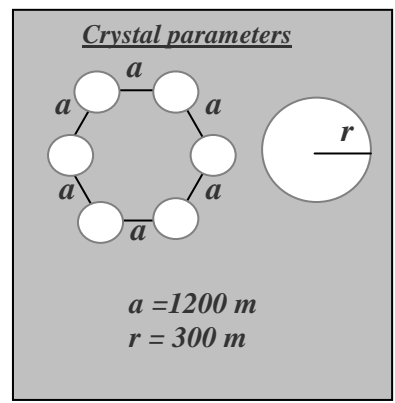

**c.** Reduction rate for honeycomb lattice configuration and circular scatter

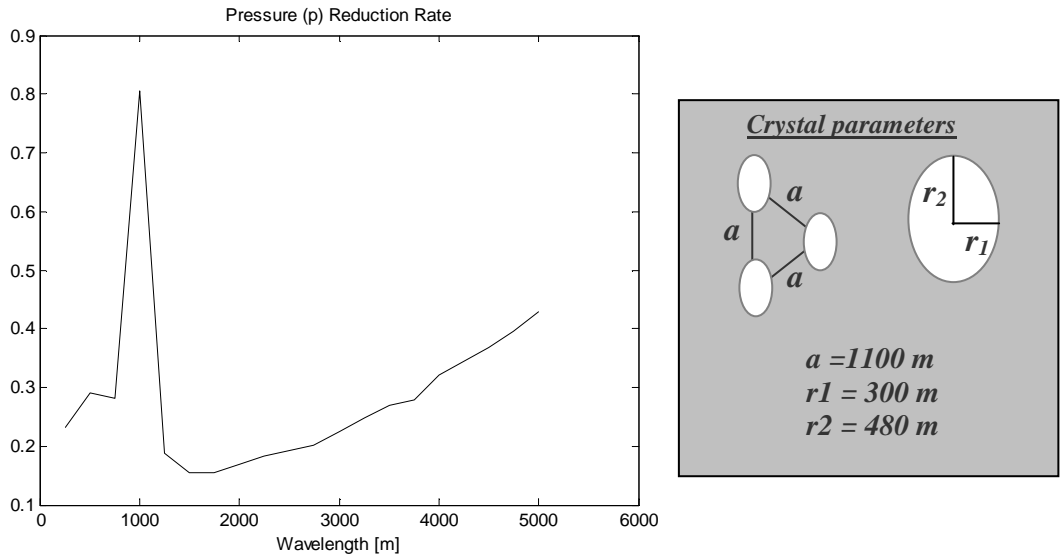

**d.** Reduction rate for triangular lattice configuration and elliptic scatter

**Figure 13**. Maximum pressure reduction rates for earthquake shielding with various crystals.

In the Figure 13.b, 13.c and 13.d, maximum pressure reduction rates versus wavelength ($\lambda$) were drawn for various seismic crystals. According to figures, seismic crystal provides considerable reduction at wavelengths between nearly 1500 m and 3000 m. One can state that these crystals could provide better reduction for wavelengths satisfying condition of $\frac{3 \cdot a}{2} < \lambda < 3 \cdot a$. In order to provide the best damping of vibration, lets locate $\lambda_{max}$ at the middle of the range of $\left(\frac{3 \cdot a}{2}, 3 \cdot a\right)$. In this case, $\lambda_{max}$ can be written as $\lambda_{max} = \frac{9 \cdot a}{4}$. According this equation, design criteria for lattice constant parameter can be written as

$$a = \frac{4 \cdot \lambda_{max}}{9} \qquad (8)$$

We see that lattice constant selection criteria given by equation (8) was valid for all seismic crystals seen in Figure 13.b, 13.c, 13.d. and it is seen that elliptic scatters exhibited better overall reduction performance than circular scatters.

## Earthquake Shield Design Concepts:

In the previous section, we theoretically demonstrated that relevant seismic crystals could damp down vibration on the surface of ground. Now, we will introduce some earthquake shielding concepts for isolation of the large areas such as cities, valleys, coasts..etc and we will discuss design problems at some decree. In the Figure 14, an application of seismic crystal for the propose of earthquake shielding was illustrated in conceptual manner. Design objective is to reduce destructive power of earthquake in a valley with soft ground. Soft ground refers area of land composed of sands, gravels and small stones. Soft ground causes swelling of ground much higher than hard ground composed of massive rocks. It is well known that earthquake was the most destructive in soft land than the rocky hills. Seismic crystal would better to build on soft ground because it will work more efficient at these lands. Building it on a geological fault will weaken the faulty plate of ground and it may trigger a earthquakes. This point should be paid attention in land selection.

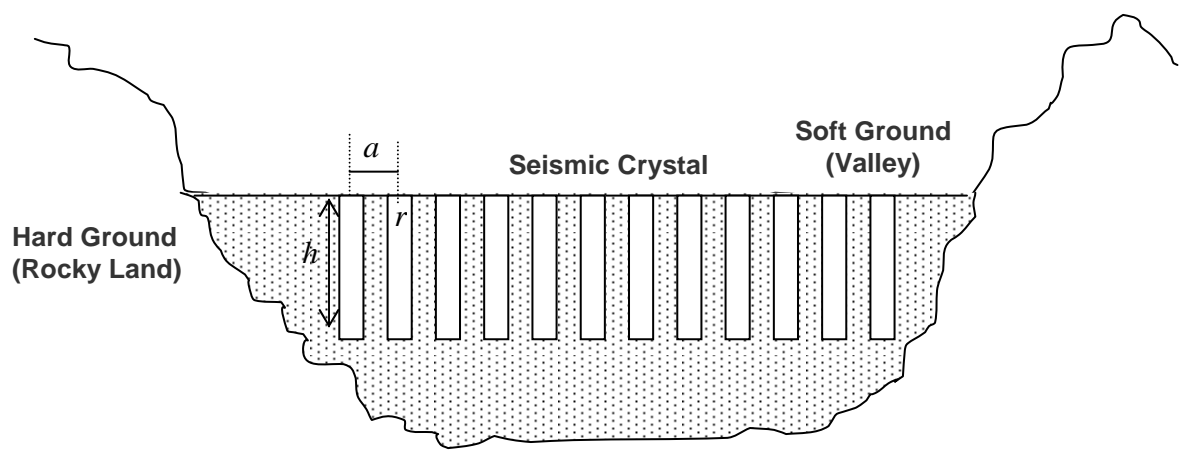

**Figure 14**. Cross-section view of earthquake shield concept for a valley

    In the figure, array of vertical holes represents scatters of seismic crystals and soft ground is host material to wave propagation. As we see in drawing, three design parameters come up, which are $r$ radius of scatters, $a$ lattice constant of crystal and $h$ height of scatters.

    In similar manner to ground, a seismic crystal can be used for reduction of tsunami wave, which occurs after earthquake in the oceans. Tsunami shield can be made of tank (bag) arrays connected each other and see bottoms. In this type seismic crystal, tank filled with air and some weights will be scatters and sea will be host material of seismic crystal. A drawing of concept was given in the Figure 15. In order to reduce negative effects of tsunami shield on sea transportation, scatter materials (Tank or bags) can be placed enough bellow of sea surface and it rise to up by filling air to tanks or bags whenever tsunami was detected. (See Figure 15.b) It may works similar to air bags system in cars.

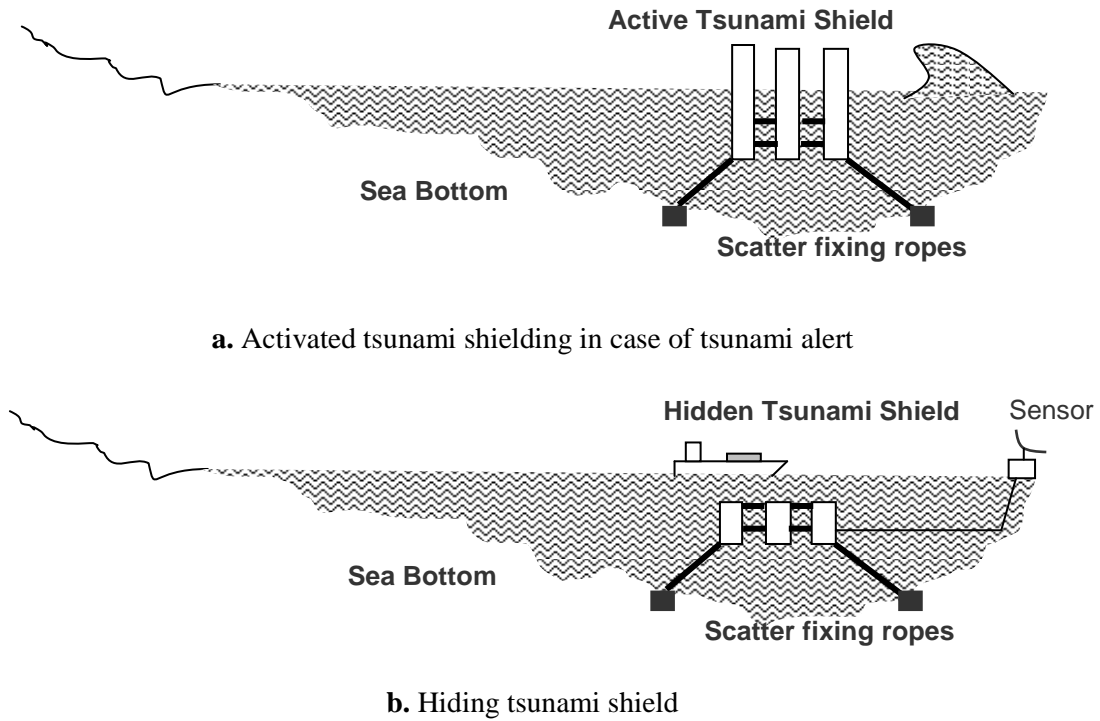

**a.** Activated tsunami shielding in case of tsunami alert

**b.** Hiding tsunami shield

**Figure 15.** Tsunami shield concept by large air-bags

Some Design Considerations;

- *Material Selection Of Seismic Crystal:* Since enormous size of earthquake shields is considered, material selection plays substantial role on feasibility. In our design concepts, we consider this point and we try to select materials, which are naturally available. So, as host materials, we prefer soft ground for earthquake shielding and water for tsunami shielding. As scatters, holes in grounds were preferred for earthquake shielding and bags to fill air for tsunami shielding.

- *Determining Design Parameters:* Design parameters, which are $r$ radius of scatters, $a$ lattice constant of crystal and $h$ height of scatters, were illustrated in the Figure 14. Determination of these parameters is not only depending on wavelength of seismic wave but also strength of crystals and cost factor. Although increasing filling fraction of crystal physically widens band gaps and reduces vibration much more, narrowing ground between scatters makes earthquake shielding fragile against seismic waves as well as increasing construction cost. In order to give idea about size of crystal, lets roughly estimate design parameters. (Better selection of parameter should be done according realistic simulations and experimental studies) Wavelength of surface wave on ground is in range of 1500-5000 m. Lattice constant ($a$) can be selected roughly half of central wavelength ($\lambda_{max}$) which has greatest amplitude in spectrum of seismic vibration. ($a = \dfrac{4 \cdot \lambda_{max}}{9}$) Radius of scatters ($r$) can be selected as $\dfrac{a}{5} < r < \dfrac{a}{3}$. Height ($h$) of scatters should be selected several times the wavelength. We summarize in the Table 1 our design parameter predictions, which is in need of validation by more advance theoretical and experimental studies.

**Table 1.** Design parameters for earthquake shielding for $\lambda_{max} = 3000$ m

| Design Parameters | Predictions for Parameter ranges | Typical Values |
|---|---|---|
| Lattice constant $a$ | $a \cong 0.44 \lambda_{max}$ | 1320 m |
| Radius of scatters ($r$) | $a/5 < r < a/3$ | 264-440 m ($a$ =1320 m) |
| Height of scatters $h$ | $h \geq 2\lambda_{max}$ | 6000 m |

- *Robustness (Strength) of Seismic Crystals:* In construction of seismic crystal, formation of lattice and body of scatters should be enough strong to resist powerful seismic waves. In order to increase strength, scatters number can be increased to enlarge width of crystal structure or conical implementation of scatters can make crystal stronger than cylindrical implementation.

Conical implementation of the scatters is drawn in the Figure 16. In addition to make scatter stronger, conical scatter usage can provide construction of seismic crystal with higher filling fractions ($f$). Conical implementation of scatters should be researched by 3D simulation model in future studies of seismic crystals.

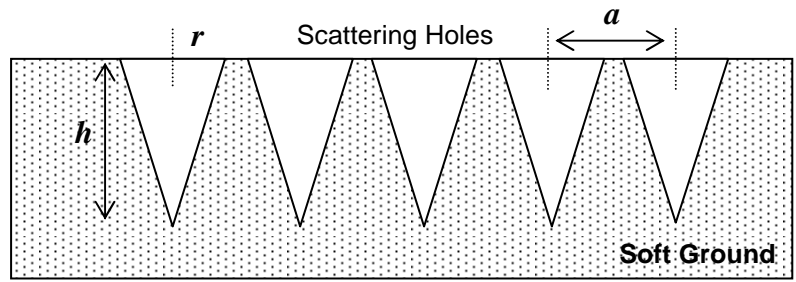

**Figure 16**. 2-D seismic crystal with conical hole scatters

## Conclusions And Future Studies

In the paper, seismic crystals were introduced by analogy with photonic and sonic crystals. Its possible applications of the earthquake shielding were theoretically demonstrated by mean of FDTD based simulations.

For the future study, more detailed research should be done on following subject;

- Theoretical study should be done by more realistic models. 3-dimensional (3D) simulation of seismic crystals should be done by parameters of real lands and recorded data set from real earthquakes.
- A study should be done for discovering more efficient crystal structure for seismic waves. Lattice geometries, scatters geometries and design parameters should be optimized to obtain higher isolation performance.
- Experimental study must be done on scaled models and its performance should also be demonstrated experimentally.

# Program Routine For 2-Dimention FDTD Based Simulations

We used elastic wave equation defined by equation (1) and (2) for FDTD based simulations. Main routine written for discrete solution applying central difference approximation to differential equations (1) and (2) was given in Program 1.

**Program 1.** Main routine of 2D FDTD based simulation

```
% q(i,j) : Normalized density matrix
% K(i,j) : Normalized bulk modulus matrix
% Rx = Δu/Δx and Ry = Δu/Δy

for n=1:nsim

% Particle velocity (Vx,Vy) updates in 2D
for i=1:W
   for j=1:H
      Vx(i,j)=Vx(i,j)-q(i,j)*Rx*(P(i+1,j)-P(i,j));
      Vy(i,j)=Vy(i,j)-q(i,j)*Ry*(P(i,j+1)-P(i,j));
   end
end

% Pressure (P) updates in 2D
for i=1:W
   for j=1:H
        P(i,j)=P(i,j)-K(i,j)*Rx*(Vx(i,j)-Vx(i-1,j))-
K(i,j)*Ry*(Vy(i,j)-Vy(i,j-1));
   end
end
end
```

**Reference:**

[1] T. Miyashita, "Sonic crystals and sonic wave-guides", Measurement Science And Technology 16, 47–63 ( 2005)

[2] X.D. Zhanga, Z.Y. Liu, "Negative refraction of acoustic waves in two-dimensional phononic crystals", Appl. Phys. Lett. 85, 341 (2004)

[3] M.M. Sigalas, "Theoretical study of three dimensional elastic band gaps with the finite-difference time-domain method", J. Appl. Phys. 87, 3122-3125 (2000)

[4] T. Miyashita, C. Inoue, "Numerical investigations of transmission and waveguide properties of sonic crystals by finite-difference time-domain method", Jpn. J. Appl. Phys. 40, 3488-3492 ( 2001)

[5] M.S. Kushwaha, B. Djafari-Rouhani "Sonic-Stop Bands For Periyodic Arrays Of Metallics Rods: Honeycomb Structure", Journal of Sound and Vibration 218(4), 697-709 (1998)

[6] E.N. Economou, M.M. Sigalas, "Classical wave propagation in periodic structures: Cermet versus network topology", Phys. Rev. B 48, 13434–8 (1993)
[7] Q.C. Zhang, X. Liu, "Far-field imaging of acoustic waves by a two-dimensional sonic crystal", Phys Reviev B 71, 054302 (2005)
[8] L.Y. Wu, L.W. Chen, "The dispersion characteristics of sonic crystals consisting of elliptic cylinders", J. Phys. D: Appl. Phys. 40, 7579–7583 (2007)